\documentclass[journal]{IEEEtran}
\usepackage{amsmath}
\usepackage{amsfonts,amssymb}
\usepackage{graphicx}
\allowdisplaybreaks[4]
\usepackage{xcolor}
\definecolor{deepred}{RGB}{205,38,38}
\usepackage{algorithm}
\usepackage{algorithmic}
\usepackage{amsthm}

\usepackage{lineno,hyperref}
\modulolinenumbers[5]
\usepackage{bm}

\usepackage{multirow}
\usepackage{array}
\usepackage{booktabs}
\usepackage{lipsum}
\usepackage{xcolor}
\usepackage{etoolbox}
\usepackage{cite}

\ifCLASSINFOpdf

\else

\fi

\ifCLASSOPTIONcompsoc
  \usepackage[caption=false,font=normalsize,labelfont=sf,textfont=sf]{subfig}
\else
  \usepackage[caption=false,font=footnotesize]{subfig}
\fi

\begin{document}
\bibliographystyle{ieeetr} 

% \title{\textcolor{blue}{Two-layer Volt/VAr Control  for Unbalanced Distribution Networks Based on MPC and Projected Newton Method}}
\title{Data-Driven Affinely Adjustable Robust \\Volt/VAr Control}
\author{\IEEEauthorblockN{Naihao Shi, \textit{Graduate Student Member}, \textit{IEEE}, Rui Cheng, \textit{Graduate Student Member}, \textit{IEEE}, Liming Liu, \textit{Graduate Student Member}, \textit{IEEE}, Zhaoyu Wang, \textit{Senior Member}, \textit{IEEE}, Qianzhi Zhang, \textit{Member}, \textit{IEEE}
% \IEEEauthorblockA{\textit{Department of Electrical and Computer Engineering} \\
% \textit{Iowa State University}\\
% Ames, IA \\
% yuanyx@iastate.edu}
}
%\thanks{This work was supported in part by the U.S. Department of Energy Wind Energy Technologies Office under Grant DE-EE0008956, and in part by the National Science Foundation under ECCS 1929975 (Corresponding author: Zhaoyu Wang).}
\thanks{Naihao Shi, Rui Cheng, Liming Liu, Zhaoyu Wang and Qianzhi Zhang are with the Department of Electrical and Computer Engineering, Iowa State University, Ames, IA 50011 USA (e-mail: snh0812@iastate.edu; ruicheng@iastate.edu; limingl@iastaet; wzy@iastate.edu; qianzhi@iastate.edu).}
% \thanks{Yifei Guo is with the Electrical and Electronic Engineering Department, Imperial College London, London SW7 2AZ, U.K. (e-mail: yifei.guo@imperial.ac.uk)}
}
\maketitle
\begin{abstract} 
This paper proposes a data-driven affinely adjustable robust Volt/VAr control (AARVVC) scheme, which modulates the smart inverter reactive power in an affine function of its active power, based on the voltage sensitivities with respect to real/reactive power injections. To achieve a fast and accurate estimation of voltage sensitivities, we propose a data-driven method based on deep neural network (DNN), together with a rule-based bus-selection process using the bidirectional search method. Our method only uses the operating statuses of selected buses as inputs to DNN, thus significantly improving the training efficiency and reducing information redundancy. Finally, a distributed consensus-based solution, based on the alternating direction method of multipliers (ADMM), for the AARVVC is applied to decide the inverter’s reactive power adjustment rule with respect to its active power. Only limited information exchange is required between each local agent and the central agent to obtain the slope of the reactive power adjustment rule, and there is no need for the central agent to solve any (sub)optimization problems.
%Under this distributed consensus-based AARVVC, the inverter’s reactive power equation slope is determined through
Numerical results on the modified IEEE-123 bus system validate the effectiveness and superiority of the proposed data-driven AARVVC method.
\end{abstract}


\section{Introduction}
\IEEEPARstart{V}{OLT}/VAr control (VVC) has always been a critical issue for power system operations. According to the standard by American National Standards Institute \cite{ANSI}, the voltage level should be maintained within a secure range, otherwise the performance of electrical equipment might be affected. Along with the growing trend of distributed energy resources (DERs), the ability of voltage support for distribution networks also needs further improvements. According to the IEEE standard 1547-2018, proactive voltage regulations are mandatory rather than optional for power systems\cite{1547}. But considering the long reaction time and high operation cost, the legacy voltage regulation devices cannot provide dynamic voltage support in shorter time periods against the fluctuating voltage issues. Compared with switch-based legacy voltage regulation devices, power electronics-based smart inverters have a much shorter response time and better controllability \cite{inverter}. They can both absorb or inject reactive power to eliminate the rapid voltage fluctuations across power systems. Authors in \cite{VVC-Challenge} declaim that the high penetration of DERs may bring more difficulties in coordinating different voltage regulation devices. 

In order to coordinate both the switch-based discrete devices and responsive smart inverters for voltage regulation, VVC problems in distribution networks are often formulated as optimal power flow (OPF) problems to maintain the system voltage level within a pre-defined range while accomplishing different objectives, e.g., minimizing system loss~\cite{loss}, reducing system cost~\cite{cost} or minimizing system voltage deviations~\cite{Vdevi}. Taking full advantage of measurements, communications and control capabilities, different VVC strategies are proposed. In \cite{central1}, a centralized  VVC framework is proposed for day-ahead scheduling of different voltage regulation devices. To address voltage issues in different timescales caused by the stochastic and intermittent nature of DER, a robust two-stage VVC strategy is proposed in \cite{2stage1} to coordinate the discrete and continuous voltage regulation devices and find a robust optimal solution, which can cope with any possible
realization within the uncertain DER output. However, the VVC problems in \cite{central1,2stage1} are solved in a centralized manner, leading to high communication costs and computational burdens. As discussed in \cite{dissurvey}, the advantages of distributed algorithms over centralized approaches in power systems include: (1) Limited information sharing, which can improve cybersecurity and protect data privacy; (2) Robustness with respect to the failure of individual agents; (3) The ability to perform parallel computations and better scalability.  
Distributed VVC strategies, based on the Alternating Direction Method of Multipliers (ADMM) \cite{admm} or projected Newton method, are applied to coordinate photovoltaic inverters \cite{distri2,Online}, and wind turbines \cite{Wind}, relying on the communication between neighboring buses/zones or the communication between the central agent and local agents.
%
% electric vehicle charging photovaltic inverters 
% Authors in \cite{distri2} propose a robust VVC strategy to minimize system loss which also divides the whole system into several sub-system. The problem is solved using the Alternating Direction Method of Multipliers (ADMM)\cite{admm}, and only communication between adjacent sub-systems is needed for the iteration process. 

% A two-level VVC strategy is introduced in \cite{two-layer1}, with a 15-min upper-level reactive power control period and a real-time droop-control based reactive power adjustments. The day-ahead switch-based devices are included in a three-stage control strategy proposed by authors of \cite{3stage}. The other two stages of the control strategy then optimizes and adjusts the inverter reactive power, respectively and a local droop control is used for adjustment. 

In the centralized and distributed VVC strategies, the reactive power outputs of DERs highly rely on communication and coordination across distribution systems, lacking the self-regulation ability of local DERs to some extent. In order to enhance the self-regulation ability of local DERs, some local voltage control strategies are proposed to combine with the centralized and distributed VVC strategies.
For instance, local voltage controls are combined with centralized/distributed VVC strategies in \cite{two-layer1,3stage,salish}. 
As one of the most common and popular local voltage control methods, droop control adjusts the reactive power outputs as a function of voltage magnitude following a given ‘Volt-Q’ piecewise linear characteristic. However, according to \cite{droop1, droop2}, the droop control may lead to some stability or feasibility issues under certain circumstances. An automatic self-adaptive local voltage control is proposed in \cite{localVVC} to improve the stability and feasibility performance, where each bus agent can locally and dynamically adjust its voltage droop function in accordance with time-varying system change.
In some works, the smart inverter's reactive power adjustment is based on its local real-time active power rather than its voltage magnitude, which can be regarded as a `P-Q' rule. More specifically, the smart inverter reactive power is adjusted as a function of its active power following a given/pre-defined ‘P-Q’ characteristic.
In \cite{QPquad}, the reactive power outputs of DERs are adjusted based on a quadratic relationship with the active power outputs. Researchers in \cite{QPdynamic} introduce a dynamic VVC strategy with several states, where the `Volt-Q' rule and the `P-Q' rule are applied to different operating statuses, respectively.

How to determine a `P-Q' rule is the key to achieving good voltage regulation performances. By projecting the complex power flow relationship into linear space, the voltage deviations caused by the power injection fluctuations can be approximated rapidly \cite{sensi} using voltage sensitivities. Taking advantage of voltage sensitivity analysis, different `P-Q' control rules for voltage regulation are  investigated. For example, in \cite{Rabih1}, an affine `P-Q' rule is introduced against the voltage deviations caused by PV uncertainties, where the reactive power adjustment ratio is obtained by solving an optimization problem with voltage sensitivities as parameters. Besides, the affine `P-Q' rule is further refined by incorporating voltage and inverter limit constraints in \cite{AARCV}, resulting in fewer voltage violations and reactive power usages. But the `P-Q' rules in \cite{Rabih1,AARCV} are determined in a system-wise centralized manner. 
In \cite{sensi:PO}, a network partition method is applied to divide the system into several zones, where the `P-Q' rule for each zone is separately determined. That is, the `P-Q' rule is determined in a zone-wise centralized manner without considering the interactions among zones. Both the system-wise and zone-wise centralized manner require a large amount of information exchanging  and computational burdens. Moreover, as mentioned before, voltage sensitivities are the key parameters for performing `P-Q' rules.
In\cite{Rabih1}, the voltage sensitivities are calculated by inverting the Jacobian matrix, requiring a large amount of computation. Authors in \cite{AARCV} utilize the surface fitting technique \cite{DataSensi}, a non-linear regression method, to estimate voltage sensitivities, where each bus voltage sensitivity is approximately calculated based on the mapping from its local power injections to its local voltage. However, this technique does not consider the influences from other buses on the local bus voltage sensitivity. The sensitivity analysis in \cite{sensi:PO} relies on the perturb and observe method, which means to repeatedly inject a small amount of power at one node and calculate the impact on bus voltages. The perturb and observe method requires repeatedly solving the power flow. 

To this end, a data-driven affinely adjustable robust Volt/VAr control (AARVVC) scheme is proposed to mitigate voltage issues against the PV uncertainty. In the first stage, the switch-based discrete devices and the base reactive power set points for PV inverters are determined  with the goal of minimizing the total system power losses. In the second stage, the reactive power outputs of PV inverters are further adjusted, based on a data-driven affine `P-Q' control rule, to reduce possible voltage fluctuations, where the `P-Q' rule is decided in a distributed manner. The main contributions of this work are listed as follows:

% As important parameters in the optimization problem, the voltage sensitivities are estimated using a data-driven method after a feature-selection process.

%{To this end, a a data-driven affinely adjustable robust Volt/VAr control (AARVVC) scheme is proposed to mitigate voltage issues against the PV outputs uncertainty. The first stage aims to get the optimal reactive power outputs from the smart inverters based on the forecast PV active power outputs. Then in the second stage, a P_Q relationship is applied to adjust the inverters' VAr. Based on the P_Q relationship, the reactive power is adjusted in accordance with the actual PV active power outputs in real-time in order to minimize the impact of PV's . In order to obtain voltage sensitivities which are parameters in the second-stage optimization, a data-driven estimation method with a pre-selection process is also designed. The main contributions of this paper can be given as follows:}
\begin{itemize}
    \item {A data-driven method, based on the deep neural network (DNN), is proposed to predict voltage sensitivities. Given the voltage magnitudes and power injections of pre-selected buses as inputs, the well-trained DNNs output the corresponding voltage sensitivity parameters, which are of great importance for determining the affine `P-Q' rule. It greatly improves the speed of calculating voltage sensitivities while maintaining high prediction accuracy.}
    \item {
    % In order to reduce the requirement of measurements to estimate the voltage sensitivities
    To improve the training efficiency and reduce redundant information, a feature-selection process, based on the rule-based bus selection with a Bidirectional Search (BDS) process\cite{bds}, is proposed. The operating statuses of each bus, including the bus active and reactive power injections and voltages, are regarded as one feature. Then the bus-selection problem can be converted into a feature-selection problem. The BDS process works from two directions for choosing the key buses: selecting the best feature while removing the worst feature until the pre-defined number of features are selected. By applying the BDS method, a subset of buses, selected by the BDS process, are combined with buses with PV installed to generate the buses selected for voltage sensitivity estimation. Only the operating statuses of those buses are applied to estimate the voltage sensitivities.}
    
    % \item {
    % \textcolor{red}{A distributed consensus-based AARVVC scheme, based on ADMM, is proposed to determine the affine `P-Q' rule, i.e., the inverter’s
    % reactive power equation slope, in a distributed hierarchical way, requiring fewer computational burdens and less information exchange compared to system-wise and zone-wise centralized manners. Under the distributed consensus-based AARVVC scheme, it relies on communication between the central agent and local buses, where the central agent is not required to solve any (sub)optimization problems.} 
    % }
    
    \item {
    The slope of the affine `P-Q' rule is obtained in a distributed hierarchical manner by using the ADMM algorithm. Relying on the communication between the central agent and local buses, the distributed consensus-based AARVVC requires less information exchange than the system-wise and zone-wise centralized manners. Plus, the central agent is only required to average its received local decision variables from local buses, thus reducing the computation burden of the central agent.}
\end{itemize}

The rest of the paper is organized as follows. Section \uppercase\expandafter{\romannumeral2} provides an overview of the proposed two-stage VVC strategy. The first-stage VVC strategy is formulated in Section \uppercase\expandafter{\romannumeral3}. Section \uppercase\expandafter{\romannumeral4} presents the second-stage VVC strategy, including the data-driven voltage sensitivity estimation and the distributed consensus-based AARVVC. Numerical results on the modified IEEE-123 bus system are given in Section \uppercase\expandafter{\romannumeral5} and the paper is concluded in Section \uppercase\expandafter{\romannumeral6}.
%Section \uppercase\expandafter{\romannumeral5} presents the simulation results and the paper is concluded in \uppercase\expandafter{\romannumeral6}.
\section{Two-Stage VVC Framework: Overview}
%The paper proposes a two-stage framework of control strategy to deal with the voltage issues in different timescales. The VCDs include traditional OLTC and CBs which can not change rapidly and the responsive DER inverters. The proposed two-stage framework can ensure the coordination of traditional VCDs and DER inverters to maintain the system voltage within the pre-defined range of [0.95, 1.05] p.u. while reducing the system line losses. 

The paper proposes a two-stage VVC framework. Based on the predicted information, the first stage aims to minimize the system power losses by dispatching the optimal settings of switch-based discrete devices and determining the optimal base reactive power set points for PV inverters. Considering the long reaction time of the discrete voltage control devices, the first-stage VVC has a slow-timescale. However, only relying on the forecast values, the intermittent nature of PV may cause unexpected voltage deviations. 
%The first-stage VVC is usually slow-timescale, but the second-stage VVC is fast-timescale, where the first-stage VVC solution keeps constant in the second-stage VVC. 
\begin{figure}[t]
    \centering
    \includegraphics[width=2.85in]{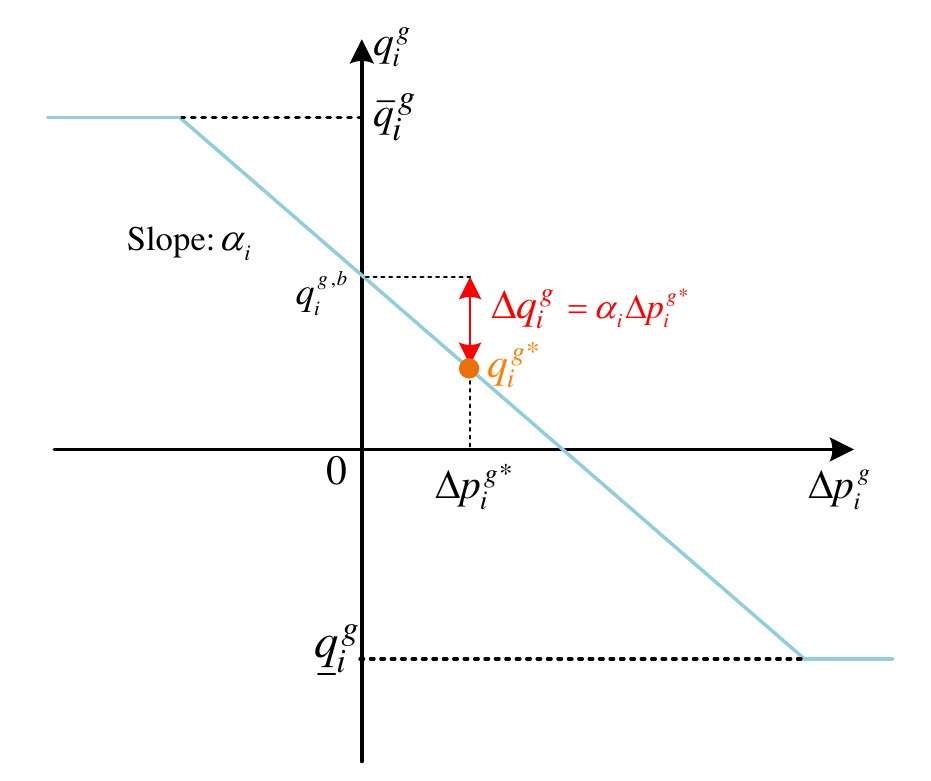}
    \caption{The reactive power adjustment following an affine 'P-Q' rule}
    \label{qslope}
\end{figure}
In the second stage, the PV deviation from its forecast value is considered. On the basis of its reactive power set point determined in the first stage, each PV inverter further adjusts its reactive power along with its real-time active power output to avoid potential voltage violations. The reactive power adjustment of PV inverter follows an optimal affine `P-Q' rule. As shown in Fig.\ref{qslope}, $q_i^{g,b}$ is the PV inverter's base reactive power set point determined in the first stage, and $\Delta{p}_i^{g*}$ is the PV deviation from its forecast value. Upon the optimal affine `P-Q' rule, the PV inverters' real-time reactive power can be adjusted as follows:
\begin{equation}\label{eq:q1}
q_i^{g*}={q}_i^{g,b}+\Delta{q}_i^g
\end{equation}
with
\begin{align}
    \Delta {{q}_i^g}={\alpha}_i\Delta{p}_i^{g*}
\end{align}
where $\alpha_i$ is the slope of the affine `P-Q' rule.

\begin{figure}[t]
    \centering
    \includegraphics[width=3.5in]{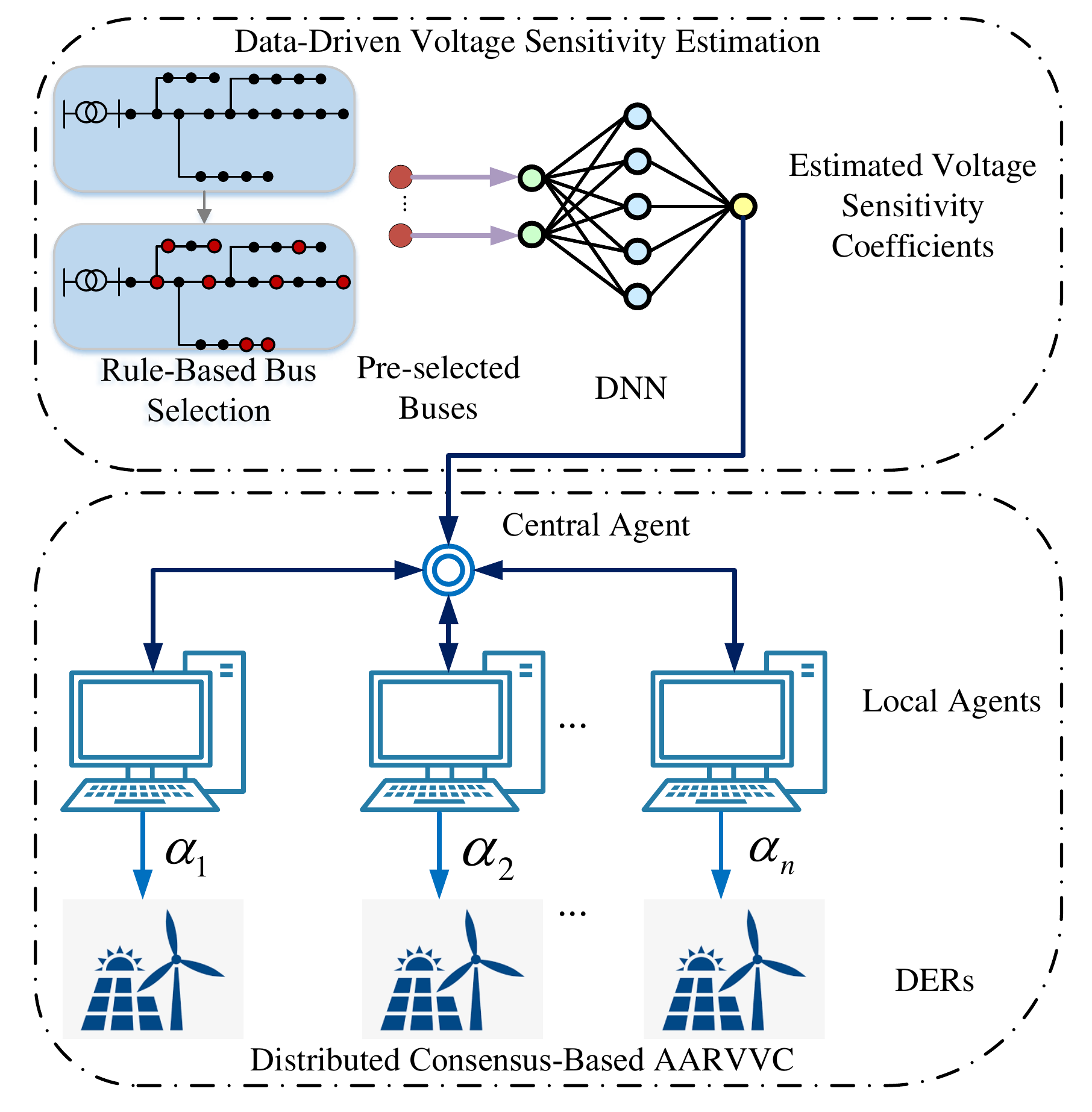}
    \caption{The data-driven AARVVC for the second-stage VVC}
    \label{flowchart}
\end{figure}

The value of $\alpha_{i}$ is determined by solving an affinely adjustable robust problem with the goal of minimizing voltage deviations caused by the PV fluctuations. Note that voltage sensitivities with respect to active/reactive power injections are the key parameters to determine the optimal affine `P-Q’ rule. Conventionally, the voltage sensitivities can be estimated by inverting the Jacobian matrix or using the perturb and observe method, which could be time-consuming. To this end, we propose a data-driven AARVVC to determine the optimal affine `P-Q’ rule in the second stage. As shown in Fig. \ref{flowchart}, the data-driven AARVVC for the second-stage VVC consists of two steps: (1) Data-driven voltage sensitivity estimation; (2) Distributed consensus-based AARVVC. 

With respect to the data-driven voltage sensitivity estimation, the DNN is utilized to predict voltage sensitivities by using the operating statuses, including the bus active and reactive power injections and voltages, as the input. The operating statuses of each bus can be regarded as one input feature for the DNN. To improve the training efficiency and reduce redundant
information behind features, a rule-based bus selection with a BDS process is first utilized to select a subset of buses whose operating statuses have a more important and greater impact on the voltage sensitivity estimation. More details about the rule-based bus selection process are provided in Section \uppercase\expandafter{\romannumeral4}. Then, the DNN-based voltage sensitivity estimation is performed to predict voltage sensitives.

Finally, a distributed consensus-based AARVVC is proposed to determine the optimal `P-Q' rule of each PV inverter in a hierarchical manner after receiving the estimated voltage sensitivities from the DNN. The communication between the local bus agents and the central agent is required for information exchange. As every local bus agent reaches a consensus with the central agent on the optimal `P-Q' rule, the communication process halts.

%In order to directly reflect the relationship of bus voltage magnitudes and power injections, sensitivity analysis is applied in our VVC strategy. Conventionally, the voltage sensitivities can be obtained by inverting the Jacobian matrix or using the perturb and observe method. To overcome the rapidly changing characteristics of renewable energy, we propose a data-driven method to estimate the voltage sensitivities to nodal active/reactive power injections based on the system operation status using deep neural networks (DNN). The estimated values of voltage sensitivities are then applied in the AARC to obtain the linear adjustment rules for the second stage control. Moreover, to reduce the measurement requirement, a feature selection process is introduced to find the key buses whose operation data are more decisive for voltage sensitivity estimation so that the accuracy can be ensured with less measurement data. 

%With the estimated voltage sensitivities from the DNN, the AARC is solved in a hierarchical distributed manner using an ADMM-based method. The AARC is split into several subproblems which can be solved by corresponding local agents. Since the second stage control aims to adjust the reactive power output of PV inverters, the local agents are located at all buses with PV installed. Once the local subproblem is solved, the solution is sent to the central agent for update. After the maximum number of iterations is achieved, the reactive power adjustment can be determined by the value of $\bm{\alpha}$ and the PV active power deviations from predicted values. 

\section{First-Stage VVC Strategy}
The first-stage VVC strategy is a deterministic OPF problem to determine the step positions of discrete devices and the optimal base reactive power set points for PV inverters based on the forecast values of DERs. The objective of this first stage is to minimize the total power losses while maintaining system voltages within the range of [0.95, 1.05]. 
\subsection{The Distribution Network}
Consider a radial distribution network containing $n+1$ buses represented as set $\{0\} \bigcup \mathcal{N}$, where $\{0\}$ denotes the slack bus at which the distribution network is connected to the transmission network and set $\mathcal{N} : =\{1,...,n\}$ denotes all other buses. Hence the radial network contains $n$ line segments connecting the adjacent buses. For any bus $j\in\mathcal{N}$, $\mathcal{N}_{j}$ is the set of all children buses of bus $j$. The set consisting all line segments in the distribution network can be expressed as: $\mathcal{L}=\{\ell_{j}=(i,j)\arrowvert i=b^{p}(j),j\in \mathcal{N}\}$, where $b^{p}(j)$ denotes the parent bus of bus $j$. For each line segment $(i,j)\in \mathcal{L}$, let $P_{ij}$ and $Q_{ij}$ represent the active/reactive power flow through the line respectively, $r_{ij}$ and $x_{ij}$ denote the line resistance and reactance. Let $p_i$ and $q_i$ represent the active and reactive power injections of bus $i$, $V_i$ and $v_i$ denote the voltage magnitude and the squared voltage magnitude of bus $i$. Then the linearized distribution power flow \cite{baran1989optimal,baran1989network} can be expressed as:
%In order to take the step-voltage regulators into consideration, a generalized branch flow model is applied. The squared voltage magnitude of bus $j$ is defined as $v_j$. For each line $(i,j)$, $t_{ij}$ denotes the transformer ratio, $n_{ij}^{tap}$ denotes the tap position between bus $i$ and bus $j$ which is an integer variable, while $\Delta tap_{ij}$ is the corresponding step of the tap changer. Note that for a line $(i,j)$ without tap changers, $t_{ij}$ is thought to be 1 and the tap position $n_{tap,ij}$ is 0. The generalized branch flow model is given as: 
%%All n branches are denoted by $\mathcal{\varepsilon}$. For every bus $j \in \mathcal{N}$, $p_{j}$ and $q_{j}$ are the bus active/reactive power injections. $\mathcal{N_{j}}$ is the set of children buses of bus $j$. For each branch connecting bus $i$ and $j$, $P_{ij}$ and $Q_{ij}$ are the active/reacitve power flow through the line while $r_{ij}$ and $x_{ij}$ are the line resistance and reactance, respectively. $v_j$ is the squard value of the voltage magnitude of bus $j$. As shown in (\ref{eq:branchflow}c), $n_{tap,ij}$ and $\Delta tap_{ij}$ denote the tap position and step. The multiplication of the quadratic term $t_{ij}^2$ and $v_i$ is included to model the voltage relationship between bus $i$ and $j$. For convexification, the quadratic term is approximately expressed in (\ref{eq:branchflow}d).
\begin{subequations}\label{eq:branchflow}
\begin{align}
    P_{ij}&=\sum_{k\in \mathcal{N}_j} P_{jk}-p_{j}\\
    Q_{ij}&=\sum_{k\in \mathcal{N}_j} Q_{jk}-q_{j}\\
    v_{i}-v_{j} &= 2(r_{ij}P_{ij}+x_{ij}Q_{ij})
\end{align}
\end{subequations}

\subsection{First-Stage VVC Problem Formulation}
% The power injections of buses are composed of the PV inverters' outputs and loads.
%\textcolor{red}{Revised by yourself first}
The active and reactive power injections of bus $j$ are denoted by $p_{j}=p^g_j-p^l_j$ and $q_{j}=q^g_j-q^l_j$, respectively. On the basis of the linearized distribution power flow, the first-stage VVC problem is formulated as:
\begin{equation}\label{eq:loss}
    \mathrm{min} \ F(t)=\sum_{(i,j)\in \mathcal{L}}r_{ij} \cdot \frac{P^2_{ij}(t)+Q^2_{ij}(t)}{v_{nom}}
\end{equation}
subject to:
\begin{subequations}\label{eq:stage1}
\begin{align}
    P_{ij}(t)&=\sum_{k\in \mathcal{N}_{j}} P_{jk}(t) + p_{j}^{l}(t) - p_{j}^{g}(t), \forall j \in \mathcal{N}\\
    Q_{ij}(t)&=\sum_{k\in \mathcal{N}_{j}} Q_{jk}(t) + q_{j}^{l}(t) - q_{j}^{g}(t)-q_{j}^{c}(t), \forall j \in \mathcal{N}\\
    v_{i}(t)-v_{j}(t)&=2(r_{ij}P_{ij}(t) + x_{ij}Q_{ij}(t)), \forall (i,j) \in \mathcal{L}\\ 
    {v}_{0}(t)&=1+2n_{tap}(t)\Delta{tap}+(n_{tap}(t)\Delta{tap})^2\nonumber\\
    &\approx 1+2n_{tap}(t)\Delta{tap}\\
    \underline{n}_{tap}&\leq n_{tap}(t) \leq \overline{n}_{tap} ,n_{tap} \in \mathbb{Z}\\
    \Big\arrowvert n_{tap}(t) &- n_{tap}(t-1) \Big\arrowvert \leq \Delta n_{tap}\\
    q_{i}^{c}(t)&=n_{i}^{c}(t) \cdot \Delta q_{i}^{c} , n_{i}^{c} \in \mathbb{Z}\\
    0&\leq n_{i}^{c}(t) \leq \overline{n}_{i}^{c}\\
    \Big\arrowvert n_{i}^{c}(t) &- n_{i}^{c}(t-1) \Big\arrowvert \leq \Delta n_{i}^{c}\\    
    -\overline{q}_{i}^{g}(t)&\leq q_{i}^{g}(t)\leq \overline{q}_{i}^{g}(t), \forall i \in \mathcal{N}\\
    \overline{q}_{i}^{g}(t) &= \sqrt{S_{i}^2-(p_{i}^{g}(t))^2}, \forall i \in \mathcal{N}\\
    \underline{v}&\leq v_{i}(t)\leq \overline{v}, \forall i \in \mathcal{N}
\end{align}
\end{subequations}
where (\ref{eq:loss}) represents the first-stage VVC goal is to minimize the total power losses. Constraints (\ref{eq:stage1}a)-(\ref{eq:stage1}c) are the linearized power flow constraints. Equation (\ref{eq:stage1}d) represents the voltage of the swing bus considering the on-load tap changing transformer (OLTC) where $n_{tap}(t)$ denotes the tap position and $\Delta{tap}$ denotes the tap step size. A linear approximation is applied to (\ref{eq:stage1}d). Equations (\ref{eq:stage1}e) and (\ref{eq:stage1}f) are the operational constraints of OLTC. The operational constraints of capacity banks and PV inverters are presented in (\ref{eq:stage1}g)-(\ref{eq:stage1}i) and (\ref{eq:stage1}j)-(\ref{eq:stage1}k), respectively. Equation (\ref{eq:stage1}l) is the  voltage constraint. By running the first-stage VVC optimization, the optimal step positions of switch-based discrete devices and the base reactive power set points
for PV inverters can be obtained. 
\section{Second-Stage VVC Strategy: Real-Time Adjustment of Reactive Power}
The second-stage VVC strategy focuses on the real-time adjustment for the reactive power outputs of inverters. In the first stage, the base reactive power set points for inverters are determined based on the forecast values of PV outputs without considering the uncertain characteristic of renewable energy. To avoid potential voltage issues caused by the PV fluctuations, the second-stage VVC is proposed for reactive power adjustment. A 'P-Q' affine rule is applied as the adjustment rule. The reactive power of PV inverter at bus $i$ after the adjustment can be expressed as (\ref{eq:q}):
%Since the PV outputs can be fluctuating and varying from the predicted values, the reactive power output adjustment is also needed in accordance with the exact values of active power output to avoid voltage deviations from the voltage profile in the first stage. 
\begin{equation}\label{eq:q}
{q}^{g*}_i={q}^{g,b}_i+{\alpha_i} \cdot {\Delta{p}^g_i}
\end{equation}
Here the PV inverter reactive power ${q_i^{g*}}$ can be split into two parts: the non-adjustable (or deterministic) part ${q^{g,b}_i}$, and the adjustable part which is expressed as an affine function of the PV deviation ${\Delta p_i^g}$ with the slope ${\alpha_i}$. Given the slope $\alpha_i$, the reactive power adjustment can be calculated immediately with the real-time PV output. Therefore, the second-stage VVC strategy allows the real-time adjustment of PV inverter's reactive power in accordance with its real-time active power output to mitigate the voltage fluctuation.
%after adjustment and $q^{g}$ is the optimal inverter reactive power set points obtained from the first stage optimization, which can be considered the non-adjustable (or deterministic) part of the inverter reactive output. As shown in equation (\ref{eq:q}), the real-time reactive power output can be represented by the non-adjustable part and also an adjustable part. By introducing the ratio $\alpha$ (\ref{eq:alpha}), a proportional relationship of the adjustment value $\Delta q$ and the deviation from forecast inverter active power output $\Delta p$ can be built. Therefore, the second stage strategy allows the real-time operation of the inverter reactive power in accordance with the renewable energy output fluctuation. 
\subsection{Second-Stage Problem Formulation: Robust Optimization Solution}
The aim of the second-stage VVC strategy is to minimize the system voltage deviations due to the rapid PV fluctuations by adjusting inverters' reactive power following the optimal affine `P-Q' rule. Let $\mathcal{N}_{G}$ denote the set of all buses with PVs installed. For any bus $i\in\mathcal{N}$, its voltage deviation can be estimated based on voltage sensitivity:
%The aim of the second-stage VVC strategy is to compute the slope $\alpha_{i}$ of the affine `P-Q' rule for each PV inverter, mitigating voltage deviations across the distribution network. Let $\mathcal{N}_{G}$ denote the set of all buses with PVs installed. Next, for any bus $i\in\mathcal{N}$, its voltage deviation can be estimated based on voltage sensitivity:
\begin{equation}\label{eq:deltaV}
    \Delta V_{i}=\sum K_{i j}^{p} \cdot \Delta p^g_j + K_{i j}^{q} \cdot \Delta q^g_j, \forall j \in \mathcal{N}_G
\end{equation}
where $K_{ij}^{p}$ and $K_{i }^{q}$ are the voltage sensitivities at bus $i$ to the active and reactive power injections at bus $j$, respectively.

It is worth mentioning that the PV deviation $\Delta{p}^g_j$ from the base PV set point ${p}^{g,b}_j$ is an uncertain parameter:
\begin{equation}\label{eq:uncertain}
    \Delta p^g_{j} \in [\Delta p _{j}^{min}, \Delta p _{j}^{max}], \forall{j}\in\mathcal{N}_G
\end{equation}
% where:
% \begin{equation}\label{eq:range}
%     \Delta p _{i}^{min}\leq 0,\quad\Delta  p _{i}^{max}\geq 0
% \end{equation}
where $\Delta p _{j}^{min}\leq 0,\quad\Delta  p _{j}^{max}\geq 0$ indicates that the actual PV outputs can deviate from the predicted values in both positive and negative directions. Considering the uncertain parameter $\Delta{p}^g_j$, the second-stage VVC problem can be formulated as a robust optimization problem:
\begin{equation}\label{eq:sumdeltaV}
\min \sum_{i=1}^{n} \Big\arrowvert \Delta V_{i} \Big\arrowvert
\end{equation}
subject to:
\begin{equation}
    (\ref{eq:deltaV}),(\ref{eq:uncertain})\nonumber
\end{equation}
% In order to simply express the relationship between nodal power injections and the voltage magnitudes, sensitivity analysis is applied which means to linearlize the non-linear voltage-power relationship at current operation point so that the impact of power variations on voltage magnitudes can be reflected directly. Considering the interaction among buses, the voltage deviations can be further expressed as:
%so that the voltage deviations caused by variations in power injection can be expressed as \ref{eq:deltaV}. After the adjustment, the voltage deviations from the original voltage set points can be expressed by: 
%$\Delta p_{j}$ reflects the active power injection deviation from the predicted value at bus $j$. Let set $\mathcal{N}_D$ be the set of all buses with DERs installed. 
% Let set $\mathcal{N}_D$ be the set of all buses with DERs installed. For $j\in \mathcal{N}_D$, $\Delta p^g_{j}$ is the forecast error of the renewable energy output at bus $j$ and $\Delta q^g_{j}$ is the corresponding reactive power adjustment at bus $j$. 
%\begin{equation}
%V_{cor,i}=\sum_{j=2}^{n} K_{i j}^{V P}\Delta p_j + K_{i j}^{V Q}\Delta q_j
%\end{equation}
% Aiming to minimize the system voltage deviations from the first stage, the optimization problem is given as: 
% \begin{equation}\label{eq:sumdeltaV}
% \min \sum_{i=2}^{n} \Big\arrowvert \Delta V_{i} \Big\arrowvert
% \end{equation}

To get rid of the absolute value operator in (\ref{eq:sumdeltaV}), an auxiliary variable $V_{i}^{aux}$ is introduced, and the problem (\ref{eq:sumdeltaV}) can  be rewritten as follows:
\begin{equation}
   \min \sum_{i=1}^{n} V_{i}^{aux}
\end{equation}
subject to:
\begin{subequations}\small\label{eq:formdeltaV}
\begin{align}
&(\ref{eq:uncertain})\nonumber\\
V_{i}^{a u x} \geq \sum_{j=1}^{n}\left(K_{i j}^{p}+\alpha_{j}\cdot K_{i j}^{q}\right)\cdot& \Delta p^g_{j},\forall{i}\in\mathcal{N},\forall{j}\in\mathcal{N}_G\\
V_{i}^{a u x} \geq \sum_{j=1}^{n}\left(K_{i j}^{p}+\alpha_{j}\cdot K_{i j}^{q}\right)\cdot& \Delta p^g_{j},\forall{i}\in\mathcal{N},\forall{j}\in\mathcal{N}_G
\end{align}
\end{subequations}

Given that $\Delta{p}^g_i$ varies in the uncertainty interval, the corresponding  affinely adjustable robust counterpart (AARC) \cite{AARC} of (\ref{eq:formdeltaV})  can be reformulated as follows:
\begin{equation}\label{eq:obj2}
    \min \sum_{i=1}^{n} V_{i}^{aux}
\end{equation}
for $\forall{i}\in\mathcal{N},\forall{j}\in\mathcal{N}_G$, subject to:
\begin{subequations}\label{eq:AARC}
\begin{align}
    &V_{i}^{a u x} \geq \sum_{j=1}^{n}\left(\theta_{i j}^{\prime} \cdot \Delta p_{j}^{max }+\theta_{i j}^{\prime \prime} \cdot \Delta p_{j}^{min }\right)\\
    &V_{i}^{a u x} \geq-\sum_{j=1}^{n}\left(\theta_{i j}^{\prime} \cdot \Delta p_{j}^{min }+\theta_{i j}^{\prime \prime} \cdot \Delta p_{j}^{max }\right)\\
    &\theta_{i j}^{\prime} \geq 0\\
    &\theta_{i j}^{\prime \prime} \leq 0\\
    &\theta_{i j}^{\prime} \geq K_{i j}^{p}+\alpha_{j} \cdot K_{i j}^{q}\\
    &\theta_{i j}^{\prime \prime} \leq K_{i j}^{p}+\alpha_{j} \cdot K_{i j}^{q}
\end{align}
\end{subequations}
where $\theta_{i j}^{\prime}$ and $\theta_{i j}^{\prime \prime}$ are the dual variables. Finally, the AARC problem reduces to a linear problem\cite{Rabih1}, whose solution is the optimal slope $\alpha_{i}$ for each PV inverter. 

With respect to the AARC problem, two main challenges should be considered:

(i) The first one is how to efficiently obtain the values of voltage sensitivities to the active/reactive power injections. Traditional methods to estimate voltage sensitivities, e.g., the inversion of Jacobian matrix and the perturb and observe method, can be time-consuming and complicated.

(ii) What's more is that the AARC problems (\ref{eq:obj2}) and (\ref{eq:AARC}) are formulated in a centralized manner, which means the central agent needs to collect all the information from local agents, leading to large computational burdens for the central agent.

To this end, we propose a data-driven AARVVC scheme consisting of the data-driven voltage sensitivity estimation and distributed
consensus-based AARVVC.
% To estimate voltage sensitivities in a more efficient way which can be more in line with the system control requirements, a data-driven method using deep neural networks (DNNs) is utilized. Aiming to reduce the measurement needs, a feature selection process is also added to select the key buses for voltage sensitivity estimation. Given the selected key buses' measurement information, the well-trained DNNs can output the estimated voltage sensitivity coefficients rapidly. Then with the sensitivity coefficients, the AARC problem is reformulated into several subproblems and solved by each local agents with a few communication with the local agent. 
%%By solving the AARC problem, the optimal ratios of the reactive power adjustments to active power deviations can be obtained. Based on the proportional ratio, once the actual output of renewable energy is available, the adjustment parts of reactive power can be simply obtained so that the voltage can maintain close to the voltage setpoints obtained in the first stage.
\subsection{Data-Driven Voltage Sensitivity Estimation}
The data-driven voltage sensitivity estimation includes the rule-based bus selection  with a BDS process and the DNN-based voltage sensitivity estimation. The rule-based bus selection with a BDS process is applied to select a subset of buses  whose operating statuses have a more important and greater impact on the voltage sensitivity estimation, thus improving the training efficiency and reducing redundant information. And the DNN-based voltage sensitivity estimation can efficiently predict voltage sensitivities with high accuracy.

\subsubsection{Rule-based bus selection with a BDS process}
% Representing the impact of network power changes on network voltages, voltage sensitivity can play an important role in the proposed control strategy for voltage regulations. 
%Two main methods: Newton-Raphson Power Flow method and Perturb-and-Observe method are widely used to obtain the voltage sensitivity. 
% A measurement-based data-driven method is applied for voltage sensitivity estimation. The dataset is based on the power flow results under different loads scenarios. 

The relationship between the voltage deviations and the deviations of bus power injections is presented as follows:
\begin{equation}\label{eq:Jacobian}
\left[\begin{array}{l}
\Delta \bm p \\
\Delta \bm q
\end{array}\right]=\bm J \cdot \left[\begin{array}{c}
\Delta \bm{\theta} \\
\Delta \bm V
\end{array}\right]
\end{equation}
where $\bm J$ is the Jacobian matrix, $\Delta \bm p$ and $\Delta \bm q$ are the deviations of bus power injections, $\Delta \bm V$ and $\Delta \bm{\theta}$  represent the deviations of voltage magnitudes and angles.
This work mainly focuses on the impact of bus power injections on voltage magnitudes. By inverting the Jacobian matrix, the relationship between the deviations of voltage magnitudes and the deviations of bus power injections can be written as:
\begin{equation}\label{eq:invJacobian}
\Delta \bm v=\left[\begin{array}{ll}
\bm{K^{p}} & \bm{K^{q}}
\end{array}\right] \cdot \left[\begin{array}{l}
\Delta \bm p \\
\Delta \bm q
\end{array}\right]
\end{equation}
where $\bm{K^{p}}$ and $\bm{K^{q}}$ in (\ref{eq:invJacobian}) are sub-matrices of $\bm J^{-1}$. The operation of matrix inversion can be time-consuming for large-scale systems.

The entries of $\bm{K^{p}}$ and $\bm{K^{q}}$ are calculated from the power flow solutions, demanding operating statuses of all buses.
However, there is always redundant information behind operating statuses of all buses. Besides, incorporating operating statuses of all buses as the input of DNN makes the training efficiency of DNN slow.

To this end, a rule-based bus selection  with a BDS Process is utilized to pick the key buses for voltage sensitivity estimation. Only the operating statuses of the selected buses will be used to perform voltage sensitivity estimation.

The operating statuses, including the bus active and reactive power injections and its voltage, of each bus are regarded as one feature, then the bus-selection problem can be converted into a feature-selection problem, which can be resolved by the BDS feature-selection method.
% Once the measurement data is available, the DNN can output the corresponding sensitivity coefficients. Using data \textcolor{blue}{from} only partial buses of the whole system, the proposed data-driven estimating method can be more practical. 

\begin{algorithm}[t]
\renewcommand{\thealgorithm}{}\selectfont
\small
\caption{\textbf{1:} BDS-Based Bus Selection}
\begin{algorithmic}
\STATE\hspace{-2mm}{\bf S1: Initialization:} Define set $F$={$\varnothing$} and set $B$=$\mathcal{N}$, $m=0$, and the number of buses to be selected $n$ .
\STATE\hspace{-2mm}{\bf S2: SFS process:}\\
Let set $\mathcal{I}$=$\{i\vert i \notin F$ and $i\in B\}$, which contains $k$ buses \{$i_{1}, i_{2},...,i_{k}$\}.\\ Initialize $i^{*}=i_{1}$, $\eta^{*}=\emph{E}(F\cup{i_{1}})$, where $\emph{E}$ is an indicator of estimation error. The larger E is, the larger the error is. \\
\textbf{for} $i=i_{1}, i_{2},...,i_{k}$,\\
\qquad $\eta$=$\emph{E}(F\cup{i})$.\\
\qquad \textbf{if} ($\eta \leq \eta^{*}$)\\
\qquad \qquad $i^{*}=i$\\
\qquad \qquad $\eta^{*}=\eta$\\
\qquad \textbf{end if}\\
\textbf{end for}\\
$F=F \cup \{i^{*}\}$\\
\STATE\hspace{-2mm}{\bf S3: SBS process:}\\
Let set $\mathcal{J}$=\{$j\vert j\notin F$ and $j\in B\}$ which contains $l$ buses \{$j_{1}, j_{2},...,j_{l}$\}.\\ Initialize $j^{*}=j_{1}$, $\mu^{*}=\emph{E}(B_{k}-{j_{1}})$.\\
\textbf{for} $j=j_{1}, j_{2},...,j_{l}$,\\
\qquad $\mu$=$\emph{E}(B_{k}-{j})$.\\
\qquad \textbf{if} ($\mu \geq \mu^{*}$)\\
\qquad \qquad $j^{*}=j$\\
\qquad \qquad $\mu^{*}=\mu$\\
\qquad \textbf{end if}\\
\textbf{end for}\\
$B=B-\{j^{*}\}$\\
\STATE\hspace{-2mm}{\bf S4: Let $m=m+1$, and go back to S2 until $m=n$, which means that the pre-defined number of buses have been selected and added to set $F$.}\\
\end{algorithmic}
\end{algorithm}

% The  process mainly relies on the Bidirectional Search (BDS) feature-selection method. The operation status (including bus voltage magnitude and power injections) of every single bus is considered one feature. 

As a sequential searching strategy, BDS consists of two separate processes: a sequential forward selection (SFS) which selects the feature that contributes most to improving the estimation accuracy from the remaining feature set, and a sequential backward selection (SBS) that deletes the feature which contributes the least to improving accuracy from the remaining feature set.

The procedure of the BDS is shown in \textbf{Algorithm 1: BDS-Based Bus Selection} in detail. In step S2, from all the remaining buses, one feature that contains the most information for voltage sensitivity estimation is selected and moved to the SFS set. In step S3, one feature that contributes the least in voltage sensitivity estimation is found and deleted from the remaining buses. Note that features selected by SFS will not be deleted by SBS while features removed by SBS will not be selected by SFS. This can ensure that the two processes can converge to the same solution from two directions. 

\begin{figure}[t]
    \centering
    \includegraphics[width=2.8in]{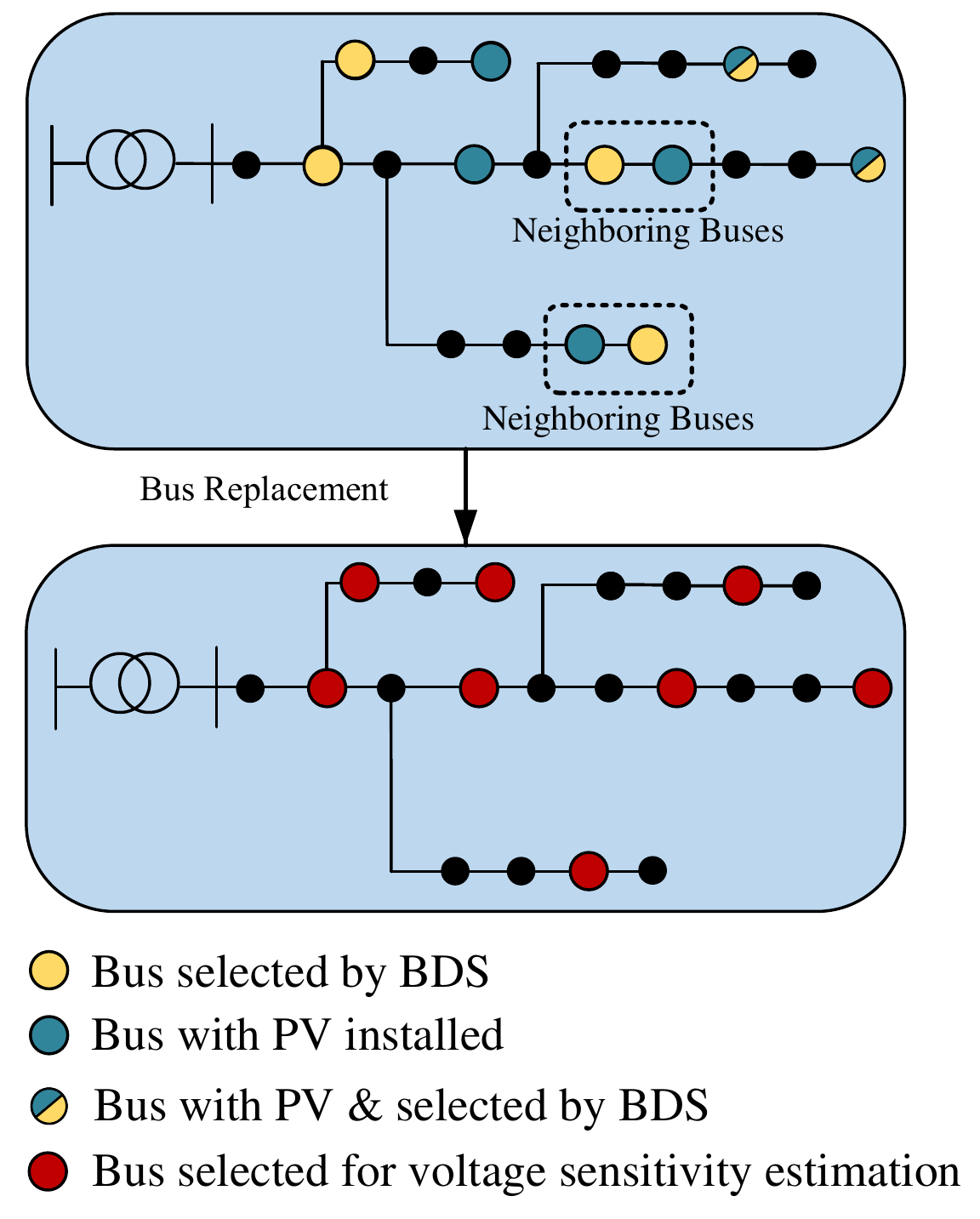}
    \caption{Merging process of buses selected by BDS and buses with PV installed}
    \label{merging}
\end{figure}

In the second-stage VVC, the PV inverter's reactive power is adjusted in accordance with its real-time active power. It indicates that the operating statuses of buses with PV installed are usually necessary for the AARVVC. From a practical point of view, to reduce the investment in measuring devices, we further define a rule to combine the key buses selected by the BDS process and the buses with PV installed. The rule is 
defined as follows: if one bus selected by the BDS process is the neighboring bus of any bus with PV installed, then the bus, selected by the BDS process, will be replaced by its neighboring bus with PV installed. This rule is based on the intuition that there are relatively strong correlations between the operating statuses of two neighboring buses. An illustration example to explain the rule to merge buses selected by BDS and buses with PV installed is depicted in Fig.\ref{merging}.

% The merging procedure does not simply add the two sets of buses together, but  includes a filter process to avoid information redundancy. The whole process is shown in. Summarizing the distributions of the selected buses and the PV inverters, there might be overlaps between the two sets and some buses are close to each other. For the overlapping buses, their operation information are certainly used for voltage sensitivity estimation. But for those neighboring buses of two sets, an evaluation process is applied, which means to test the prediction performance with and without the bus from the BDS set. If the performance is close, then the bus can be abandoned. Otherwise, the operation information of the bus should be kept for sensitivity estimation.
\subsubsection{A DNN-based voltage sensitivity estimation}
%In this work, the distribution of PVs are also taken into consideration in selecting the location of measurements. Here we assume that the bus with PV installed is already equipped with measurement. The process of the bus-selection based on both the BDS selection results and the PV bus distribution is shown in The process relies on a rule-based combination which can further reduce the number of buses to be observed while maintaining comparable accuracy. For each bus selected by the BDS method, when there's a PV bus close to it, try to use the PV bus instead for the voltage sensitivity estimation and evaluate the performance. If the performance is comparable, then this bus can be replaced by the PV bus. 

The buses, selected by the proposed rule-based bus selection, are used for voltage sensitivity estimation. Instead of requiring the operation statuses of the whole system, only the operating statuses of selected buses are set as the input of DNN. Aiming to establish the mapping relationship from the input features to the voltage sensitivities, supervised machine learning, using a three-layer fully connected DNN, is performed. With the help of the well-trained DNN, the estimated voltage sensitivities can be obtained in real-time. Compared with the conventional methods to calculate the voltage sensitivities, the DNN-based voltage sensitivity estimation can be much more efficient and more capable of coping with the rapidly changing operating statuses of power systems. 

\subsection{Distributed Consensus-Based AARVVC}
To obtain the slope of the affine 'P-Q' rule for PV inverter in a distributed manner, we propose the distributed consensus-based AARVVC to solve the AARC problem (\ref{eq:obj2})-(\ref{eq:AARC}).  For each bus $i\in\mathcal{N}$, we introduce $\bm{z}_i=\{z_i^j|z_i^j=\alpha_{j}, \forall{j}\in\mathcal{N}_G$\}, and let $\bm{z}=\{\bm{z}_i|\forall{i}\in\mathcal{N}\}$. Then the AARC problem (\ref{eq:obj2})-(\ref{eq:AARC}) can be reformulated as follows:
%Using ADMM, the overall AARC problem is divided into several subproblems by introducing the auxiliary variable $z_{i}^{j}$. The AARC problem can be rewritten as: 
\begin{equation}\label{eq:objDAARC}
    \min \sum_{i=1}^{n} V_{i}^{aux} 
\end{equation}
for $\forall{i}\in\mathcal{N},\forall{j}\in\mathcal{N}_G$, subject to:
\begin{subequations}\label{eq:DAARC}
\begin{align}
    &V_{i}^{a u x} \geq \sum_{j=1}^{n}\left(\theta_{i j}^{\prime} \cdot \Delta p_{j}^{\max }+\theta_{i j}^{\prime \prime} \cdot \Delta p_{j}^{\min }\right)\\
    &V_{i}^{a u x} \geq-\sum_{j=1}^{n}\left(\theta_{i j}^{\prime} \cdot \Delta p_{j}^{\min }+\theta_{i j}^{\prime \prime} \cdot \Delta p_{j}^{\max }\right)\\
    &\theta_{i j}^{\prime} \geq 0\\
    &\theta_{i j}^{\prime \prime} \leq 0\\
    &\theta_{i j}^{\prime} \geq K_{i j}^{p}+z_{i}^{j} * K_{i j}^{q}\\
    &\theta_{i j}^{\prime \prime} \leq K_{i j}^{p}+z_{i}^{j} * K_{i j}^{q}\\
    &z_{i}^{j}=\alpha_{j}
\end{align}
\end{subequations}
Note $\Delta \bm{p}^{min}=[\Delta p _{j}^{min}]_{j\in\mathcal{N}_G}, \Delta \bm{p}^{max}=[\Delta p _{j}^{max}]_{j\in\mathcal{N}_G}$ are the uncertain parameters, which are assumed to be accessed by each bus $i\in\mathcal{N}$ in this paper, and  $K_{i j}^{p}, K_{i j}^{p}$ are the voltage sensitivity of bus $i$ with respect to the active and reactive power of bus $j$, which can be accessed by bus $i$. It is worth mentioning $K_{i j}^{p}, K_{i j}^{p}$ can be estimated by the proposed data-driven voltage sensitivity estimation.

In addition, $\theta_{i j}^{\prime},\theta_{i j}^{\prime \prime} $ can be regarded as the variables associated with bus $i$. In this case, the objective function (\ref{eq:objDAARC}) as well as the constraints (\ref{eq:DAARC}a)-(\ref{eq:DAARC}f) can be split into subproblems related to each bus $i\in\mathcal{N}$. Then, the only coupling constraint is (\ref{eq:DAARC}g).

To deal with the coupling constraint (\ref{eq:DAARC}g), let $\bm{\lambda}=\{\bm{\lambda}_i|i\in\mathcal{N}\}$, where $\bm{\lambda}_i=\{\lambda_{i}^{j}|j\in\mathcal{N}_G\}$, denote dual variables associated with (\ref{eq:DAARC}g), then the augmented Lagrangian function can be written as:
%To reformulate the problem into a distributed manner, the auxiliary variables $z_{i}^{j}$ are introduced. Then the optimization problem can be rewritten as:
%By introducing $z_{i}^{j}$, the original optimization problem can be split into several subproblems and each subporblem is a linear program. The subproblems are solved at corresponding buses with PV installed. The augmented Lagrangian is given as: 
\begin{equation}\small\label{eq:lag}
\begin{split}
&L_{\rho}(\bm{\alpha},\bm{z},\bm{\lambda})=  \sum_{i=1}^{n}L_{\rho}^{(i)}(\alpha_{i},\bm{z}_i,\bm{\lambda}_i)
\\&=\sum_{i=1}^{n}
\left[ V_{i}^{a u x} +\sum_{j\in\mathcal{N}_G} \left(\lambda_{i}^{j} \cdot \left(z_{i}^{j}-\alpha_{j}\right)+\frac{\rho}{2}\cdot \parallel z_{i}^{j}-\alpha_{j} \parallel ^{2} \right) \right]
\end{split}
\end{equation}
where $\rho$ is a parameter. Based on ADMM, the problem (\ref{eq:objDAARC})-(\ref{eq:DAARC}) can be solved in a distributed manner, which is shown in detail in \textbf{Algorithm 2: Distributed Consensus-Based AARVVC}.

As seen in S2 and S3 of Algorithm 2, each local agent is assigned its own subproblem to obtain the optimal values of $\bm z_{i}(k)$ and then communicates $\bm z_{i}(k)$ to the central agent using the communication capacity of the inverters during the $k$-th iteration. Then in step S4, the consensus-based ADMM also simplifies the iteration process and the update of $\alpha_{j}$ can be realized by simply averaging all 
entries in the $j$th column of $\bm{z}$, and the values of $\alpha_{j}$ are then sent back to corresponding local agents. The local agents then update the dual variable $\bm{\lambda}_{i}$ based on the updated $\bm{\alpha}$, $\bm{z}_{i}$ and the parameter $\rho$ in step S5. The iteration process will stop until the consensus is achieved among all the local agents or the maximum number of iterations is reached. 

%once the maximum iteration number is achieved. 
\begin{algorithm}[t]
\renewcommand{\thealgorithm}{}\selectfont
\small
\caption{\textbf{2:} Distributed Consensus-Based AARVVC}
\begin{algorithmic}
\STATE\hspace{-2mm}{\bf S1: Initialization. Let the number of iterations $k=1, \bm{\alpha}(1)=0, \bm{z}_i(1)=0, \bm{\lambda_{i}}(1)=0$, $\rho>0$.} 
\STATE\hspace{-2mm}\bf S2: Each local bus agent $i$ updates $\bm{z}_i(k)$ based on the voltage sensitivities $K_{i j}^{p}$ and $K_{i j}^{q}$.
\begin{align*}\small
\bm{z}_i{(k+1)}&=\arg\min_{\bm{z}_i} L_{\rho}^{(i)}({\bm{\alpha}(k+1)},{\bm{z}_i},{\bm{\lambda}_i(k)})\\
&\text{s.t.~}(\ref{eq:DAARC}a) - (\ref{eq:DAARC}f)
\end{align*}
\STATE\hspace{-2mm}{\bf S3: Each local agent then communicates $\bm{z}_i(k+1)$ to the central agent.}
\STATE\hspace{-2mm}{\bf S4: Collecting $\bm{z}_{i}(k)$ from each local bus agent $i\in\mathcal{N}$, the central agent then updates  $\bm{\alpha}(k+1)$. Each entry $\alpha_j(k+1)$} of $\bm{\alpha}(k+1)$ can be expressed as:
\begin{align*}\small
\alpha_{j}{(k+1)}=\frac{\sum_{i\in\mathcal{N}}{z_{i}^{j}}(k+1)}{n+1}, \forall{i}\in\mathcal{N}, \forall{j}\in\mathcal{N}_G
\end{align*}
 The central agent then sends $\bm{\alpha}{(k+1)}$ back to each local bus agent $i$.
\STATE\hspace{-2mm}{\bf S5: Each local bus agent $i$ updates $\bm{\lambda}_{i}(k+1)$:}
\begin{align*}\small
\bm{\lambda}_{i}(k+1)=&\bm{\lambda}_{i}(k)+\rho \cdot (\bm{z}_{i}(k+1)-\bm{\alpha}(k+1)), \forall{i}\in\mathcal{N}
\end{align*}\small
\STATE\hspace{-2mm}{\bf S6: Let $k=k+1$. If $k>k_{max}$, or the consensus is achieved, stop the iteration process; otherwise, go to $\textbf{S2}$, where $k_{max}$ is the maximum number of iterations.} 
\end{algorithmic}
\end{algorithm}

\section{Numerical Results}
\label{sec:Case}
%The performance of the proposed AARVVC is tested on the modified IEEE-123 bus test system. First of all, the performance of the data-driven voltage sensitivity estimation method is tested. The dataset is produced by Matpower, a Matlab package for steady-state power system simulation. Based on the randomly generated power injections, Matpower can calculate the corresponding system operation information. The bus power injections as well as voltages are recorded as the inputs. Then by forming the Jacobian matrices and calculating the inverse, the voltage sensitivities can be calculated and regarded the outputs. After a feature-selection process, a set of key buses who can contribute more information in voltage sensitivity estimation are picked out and a DNN are trained based on information of the selected buses. Taking voltage sensitivities as important parameters, the inverters' reactive power adjustment ratios $\bm{\alpha}$ can be further obtained by solving an optimization problem in the distributed manner and applied to eliminate the voltage deviations caused by the PV uncertainty. 

\begin{figure}[t]
    \centering    \includegraphics[width=3.5in]{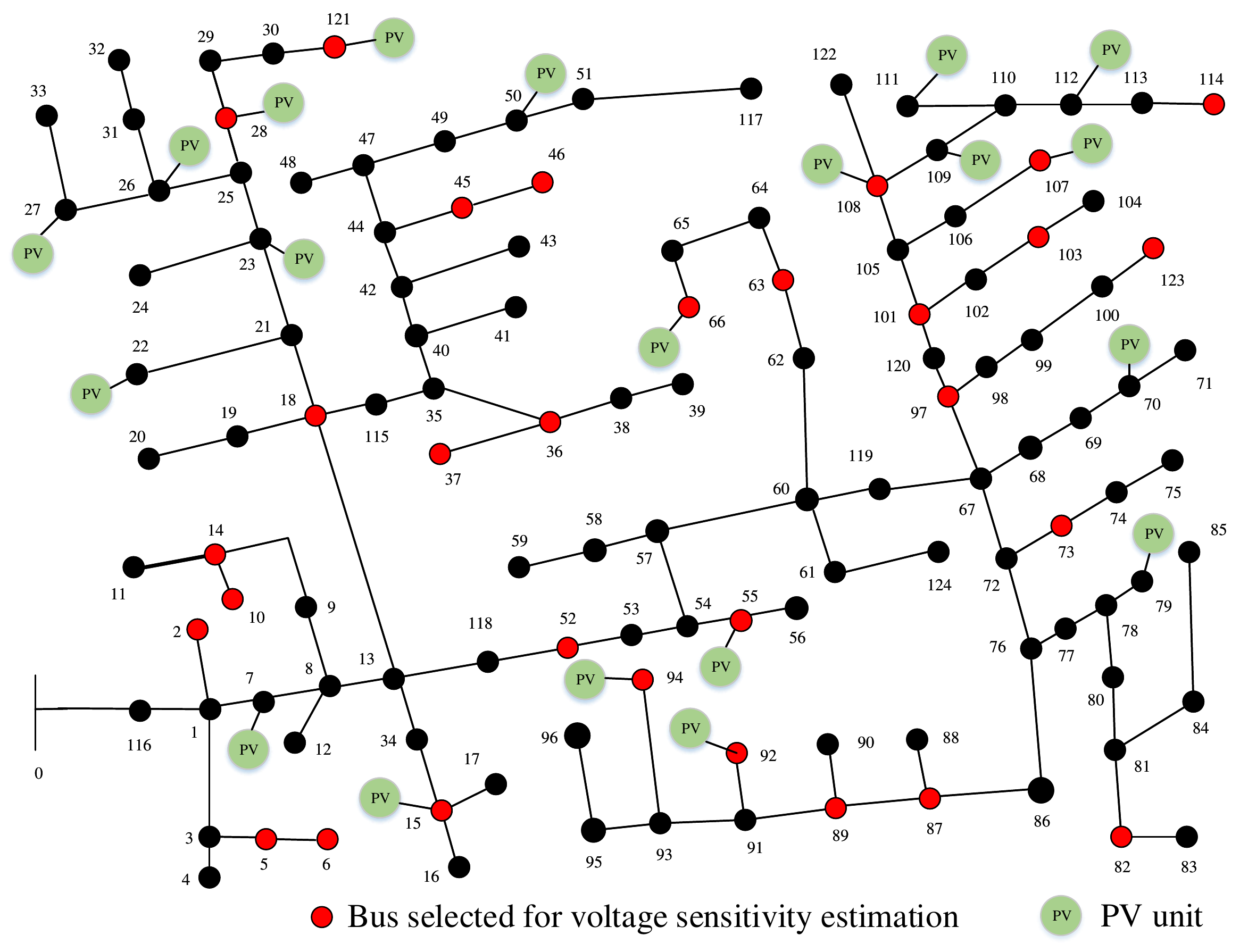}
    \caption{The modified IEEE-123 bus test system}
    \label{testsystem}
\end{figure}

In this section, the proposed data-driven AARVVC is implemented on the modified IEEE-123 bus test system to test its performance. 
The modified IEEE-123 bus test system with PV generators is shown in Fig.~\ref{testsystem}. 
The base voltage for the test system is set to 4.16 kV and the base power is set to 100 kVA. The first-stage VVC strategy is run at a circle of 15 minutes based on the forecast PV generations to dispatch the switch-based discrete devices, e.g., OLTC, and determine  the base reactive power set points for PV inverters. The forecast PV penetration of this system is 47.79\% in the firs-stage VVC. For each single PV, a 50\% uncertainty interval is considered in the second-stage VVC, indicating the uncertainty set of the PV penetration of this system can be 23.89\% to 71.68\%. 

{The step positions of discrete devices} keep unchanged within the second stage.
The reactive power of PV inverter is adjusted following the optimal affine `P-Q' rule, determined by the proposed data-driven AARVVC. In the distributed consensus-based AARVVC, the parameter $\rho$ is set as 0.01 and the maximum number of iterations is set as 100.
%The training dataset for voltage sensitivity estimation was generated using Matpower, a Matlab package for steady-state power system simulation and the distributed consensus-based AARVVC was solved in IBM ILOG CPLEX 12.10 solver. 

\subsection{Voltage Sensitivity Comparisons}
As discussed before, the data-driven voltage sensitivity estimation includes two main parts: 
the bus-selection process and the DNN-based voltage sensitivity estimation,
where the operating  statuses of these selected buses are used as the input of DNN for voltage sensitivity prediction.

\begin{figure}[t]
    \centering
    \includegraphics[width=3.5in]{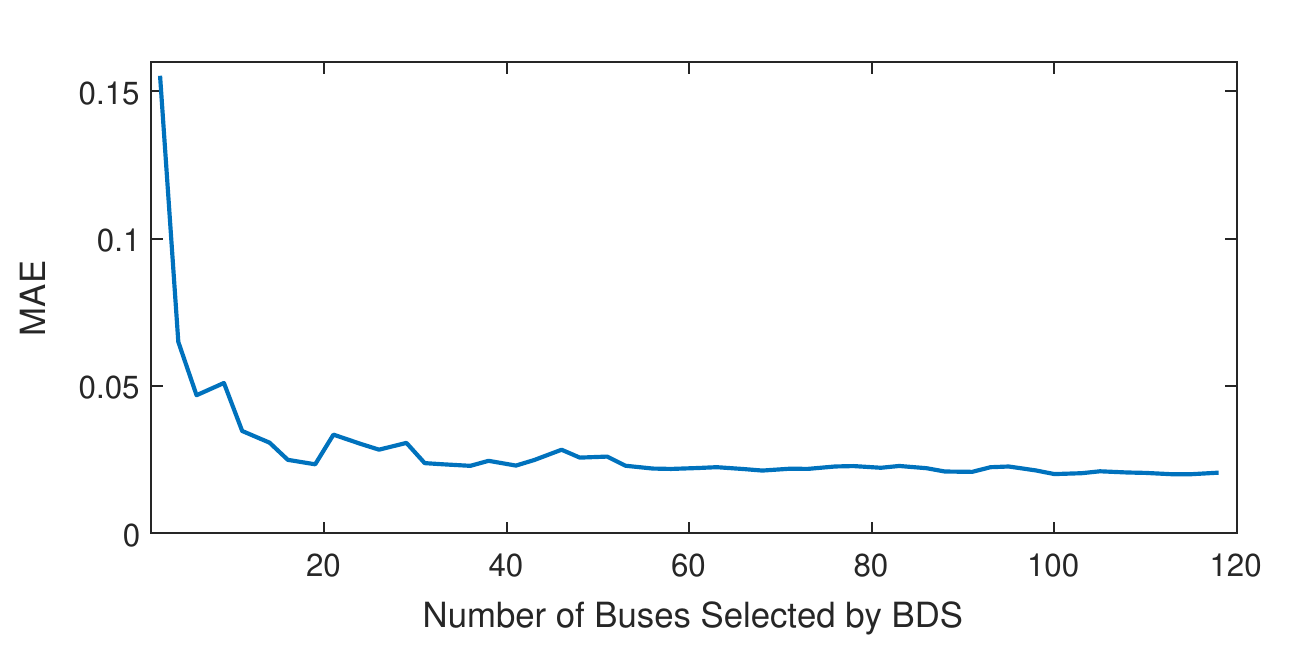}
    \caption{MAE versus the number of selected buses.}
    \label{MAE}
\end{figure}

To evaluate the impact of the number of selected buses 
on the prediction accuracy, the mean average error (MAE) is chosen as the evaluation metric, which can be expressed as follows: 
 \begin{equation}\label{eq:MAE}
  MAE=\frac{1}{n_c} \sum_{i=1}^{n_c}\left|x_{i}- \hat{x}_i\right|
\end{equation}
where $n_c$ is the number of entries of the predicted voltage sensitivities, $x_{i}$ represents the real voltage sensitivity and $\hat{x}_i$ is the estimated voltage sensitivity. 
The impact of the number of selected buses on MAE is depicted Fig.~\ref{MAE}. 
As can be seen Fig.~\ref{MAE}, MAE first decreases sharply as the number of selected buses increases, then MAE shows slight fluctuations as the number of selected buses is greater than 20. 
It shows that after the number of selected buses reaches 30, incorporating operating statuses of more buses does not contribute much to improving the prediction accuracy of voltage sensitivity. This phenomenon indicates there is redundant information behind the operating status of all the buses. 

% In this work, the number of key buses selected by the BDS method is set to 15. At the same time, 20 PV inverters are installed in the modified IEEE-123 bus test system whose reactive power can be adjusted in the AARVVC for voltage regulation. Since in the second-stage reactive power control, PV buses' operation information is already required, they are also included in voltage sensitivity estimation.
In this case, the number of selected buses to perform voltage sensitivity estimation is set to 30.  The results of the bus-selection process for the modified IEEE-123 bus test system, selected by the proposed rule-based voltage sensitivity in Section \uppercase\expandafter{\romannumeral4}, are depicted as red dots in Fig.\ref{testsystem}. 
Those selected buses are distributed across the distribution network. It indicates information coming from almost all parts of the distribution network is incorporated in those selected buses. This might shed light on the reason why using the operating status of part of buses is enough to achieve the  accurate voltage sensitivity estimation.
% By merging the key buses and the PV buses and evaluating the impact of deleting the neighboring buses, a set of 30 buses are finally determined for estimating voltage sensitivities. The results of the bus-selection process is presented in Fig.~\ref{testsystem}. The topology of the test system with the distribution of PVs are given in Fig.\ref{testsystem}. The red dots in Fig.\ref{testsystem} represent the siting of buses selected for sensitivity estimation. 
% which means that for accurately estimating the voltage sensitivities, the operation information from almost all parts of the system is required. By selecting the key buses which can contain more information, the requirements of measurements can be greatly reduced. 
% \begin{figure}[t]
%     \centering
%     \includegraphics[width=3.5in]{123bus.pdf}
%     \caption{Topology of the test system and buses selected for voltage sensitivity estimation}
%     \label{testsystem}
% \end{figure}

\begin{figure}[t]
    \centering
    \includegraphics[width=3.1in]{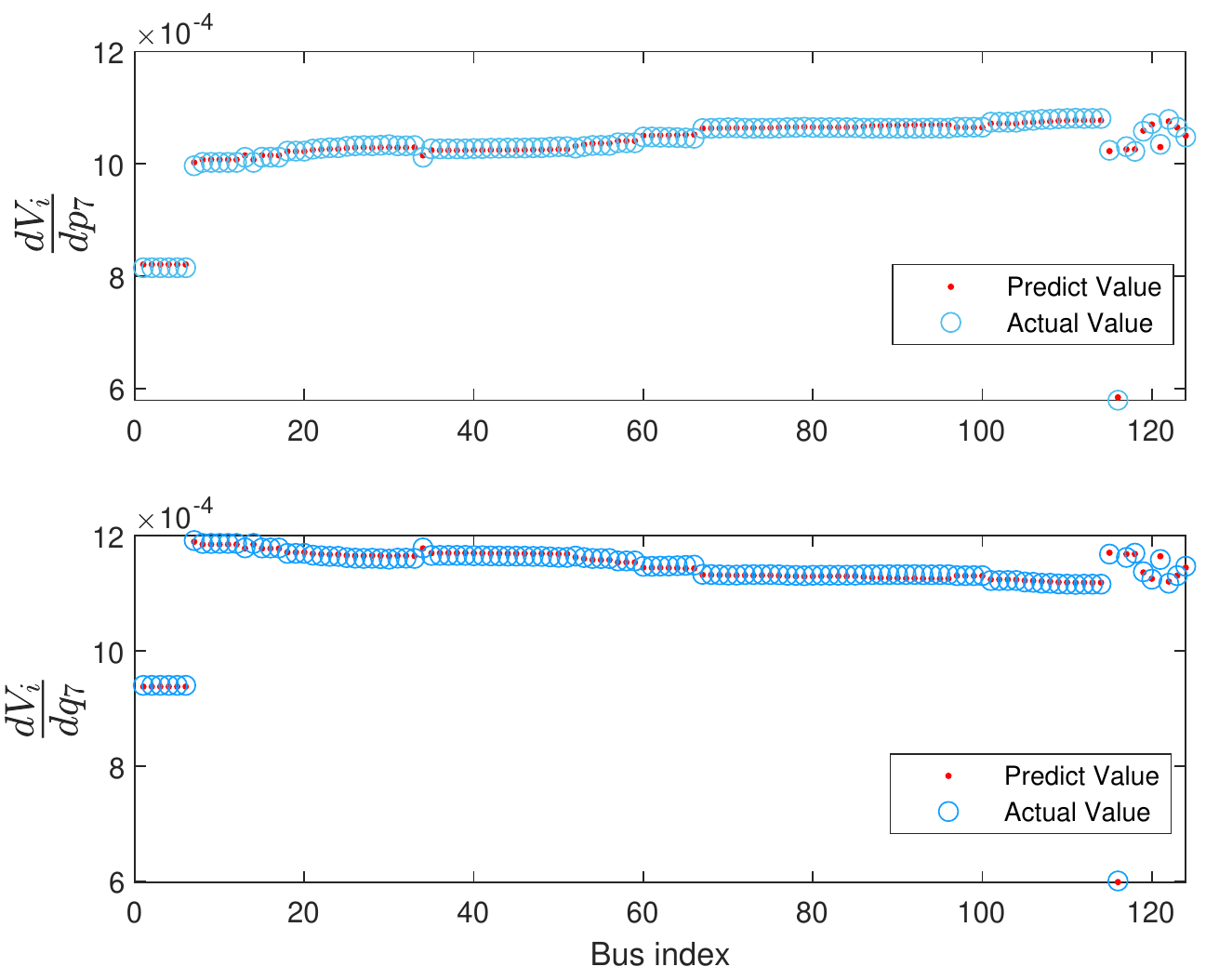}
    \caption{Actual and estimated voltage sensitivities with respect to active and reactive power injections at bus 7}
    \label{compare_pre&real}
\end{figure}

Taking bus 7 as an example, Fig. \ref{compare_pre&real} shows the actual and estimated voltage sensitivities of each bus $i\in\mathcal{N}$ with respect to the active and reactive power injection at bus 7, i.e., $dV_i/dp_7$ and $dV_i/dq_7$ for $\forall{i}\in\mathcal{N}$.  
 The actual voltage sensitives are calculated 
by inverting the Jacobian matrix, which are regarded as the benchmark, and the estimated voltage sensitivities are calculated from the proposed data-driven voltage sensitivity estimation method. As shown in  Fig. \ref{compare_pre&real}, the values of the estimated and actual voltage sensitivities are very close. It validates that the proposed data-driven voltage sensitivity estimation method  provides accurate prediction of the voltage sensitivities by only making use of the information from the selected buses.

\subsection{Performance of the Distributed Consensus-Based AARVVC}
As important parameters, the voltage sensitivities with respect to bus power injections, to decide the slope of the affine `P-Q' rule $\alpha_{i}$, it has been validated in subsection \ref{sec:Case}-A that the proposed data-driven voltage sensitivity estimation method can accurately predict voltage sensitivities. We further test the performance of our proposed Algorithm 2: Distributed Consensus-Based AAARVC.

\begin{figure}[t]
    \centering
    \includegraphics[width=3.1in]{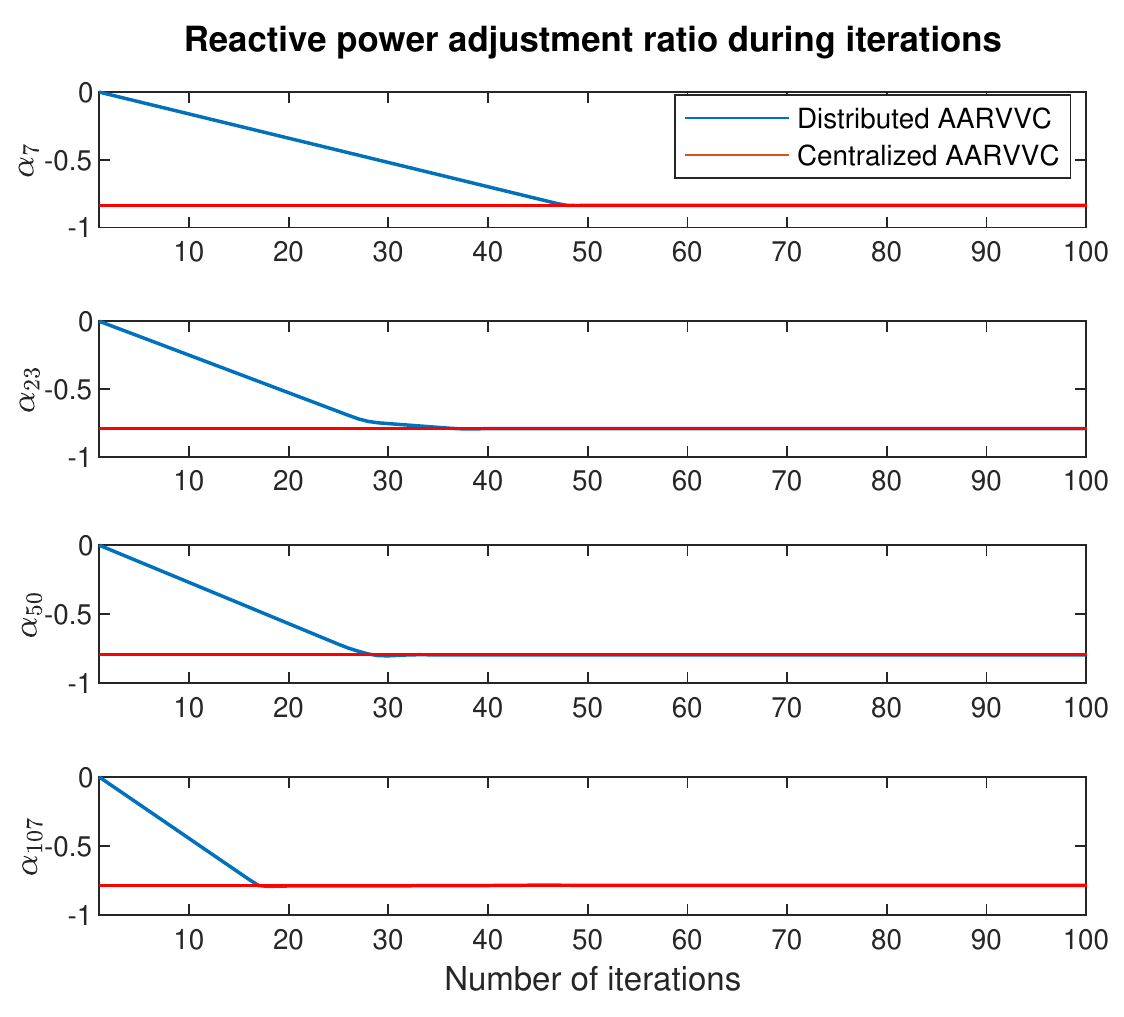}
    \caption{Slopes for PV inverters at buses 7, 23, 50, and 107}
    \label{admm_conv}
\end{figure}

Once the estimated voltage sensitivities are given, the slope $\alpha_{i}$ of the affine `P-Q' rule for each PV inverter can be determined by our proposed Algorithm 2: Distributed Consensus-Based AAARVC. 
Taking PV inverters at buses 7, 23, 50 and 107 as an example, the adjustment slopes for those PV inverters, determined by the distributed consensus-based AAARVC, 
are shown in Fig.~\ref{admm_conv}. The adjustment slopes for those PV inverters solved by the centralized optimization, i.e., the AARC problem (\ref{eq:obj2}) and (\ref{eq:AARC}) is solved in a centralized manner, are  depicted in Fig.~\ref{admm_conv} as the benchmark. It can be observed from Fig.~\ref{admm_conv} that all those slopes, determined by the distributed consensus-based AAARVC, can converge to the benchmark, the slopes determined by the centralized optimization.
It means that the optimal `P-Q' rules can be accurately calculated by our proposed distributed consensus-based AAARVC in a hierarchical distributed manner.
\subsection{Algorithm Comparisons}
For algorithm comparisons, four different VVC schemes are considered:

Scheme 1-First-stage VVC: Only the first-stage VVC is considered.

Scheme 2-Centralized AARVVC with accurate voltage sensitives: The AARC problem (\ref{eq:obj2}) and (\ref{eq:AARC}) is solved in a centralized manner, where the voltage sensitivities are obtained by inverting the Jacobian matrix.

Scheme 3-Distributed consensus-based AARVVC with  accurate voltage sensitives: The AARC problem (\ref{eq:obj2}) and (\ref{eq:AARC}) is solved in a distributed consensus-based manner, where the voltage sensitivities are obtained by inverting the Jacobian matrix.
% On the basis of (ii), but the optimal adjustment rule is determined in a hierarchical distributed manner.

Scheme 4-Our proposed data-driven AARVVC, i.e., distributed consensus-based AARVVC with estimated voltage sensitives: 
The AARC problem (\ref{eq:obj2}) and (\ref{eq:AARC}) is solved in a distributed consensus-based manner, where the voltage sensitivities are estimated by the proposed data-driven voltage sensitivity estimation method.

Note that in Scheme 1, the reactive power outputs of PV inverters are maintained at the solution of  the first-stage VVC without any adjustments. In Schemes 2-4, the reactive power outputs of PV inverters are adjusted by the determined `P-Q' affine rule.
\begin{figure}[t]
    \centering
    \includegraphics[width=3.1in]{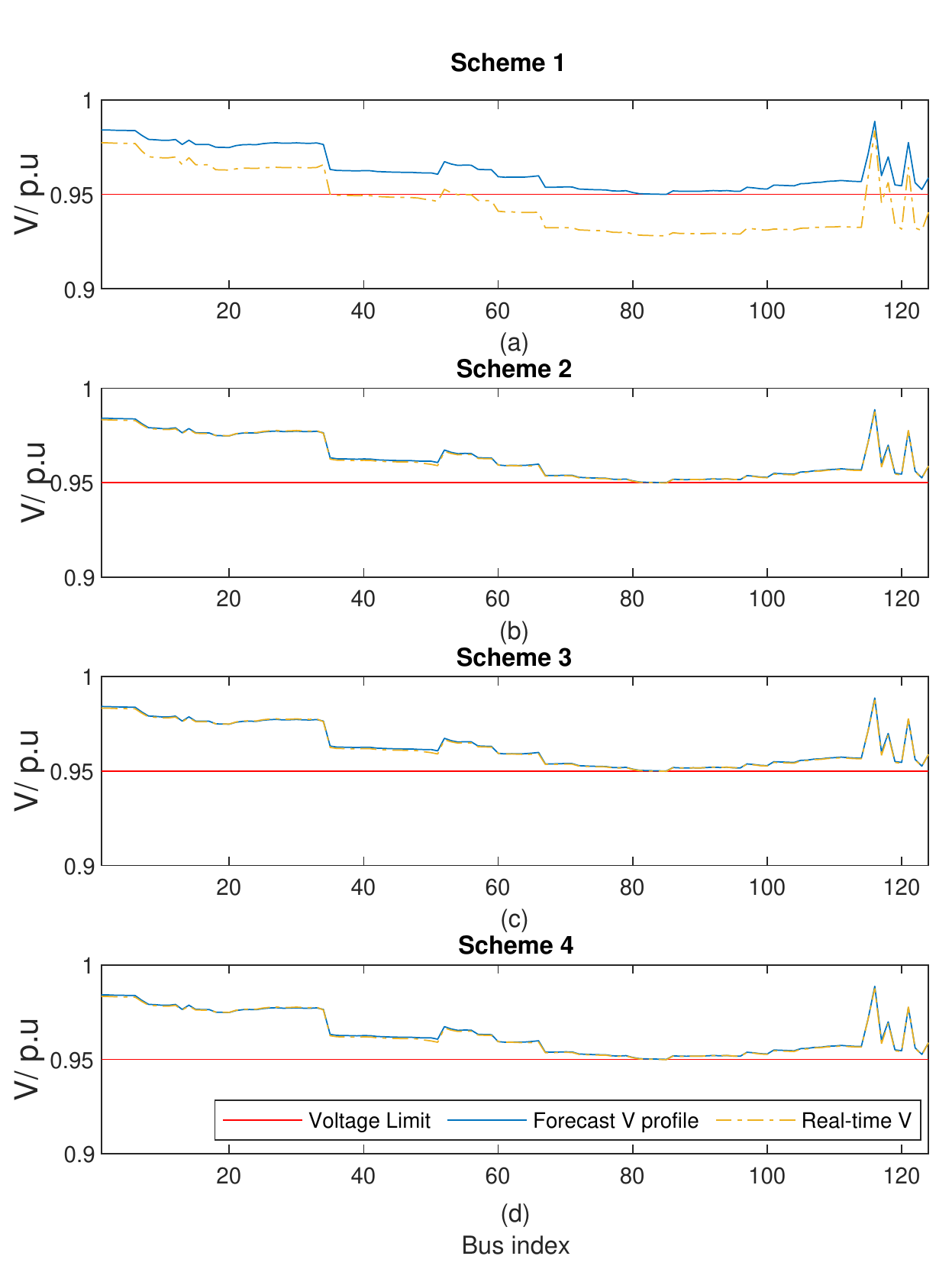}
    \caption{The voltage profiles of different schemes under an extreme scenario}
    \label{V1point}
\end{figure}

\begin{figure}[t]
    \centering
    \includegraphics[width=3.2in]{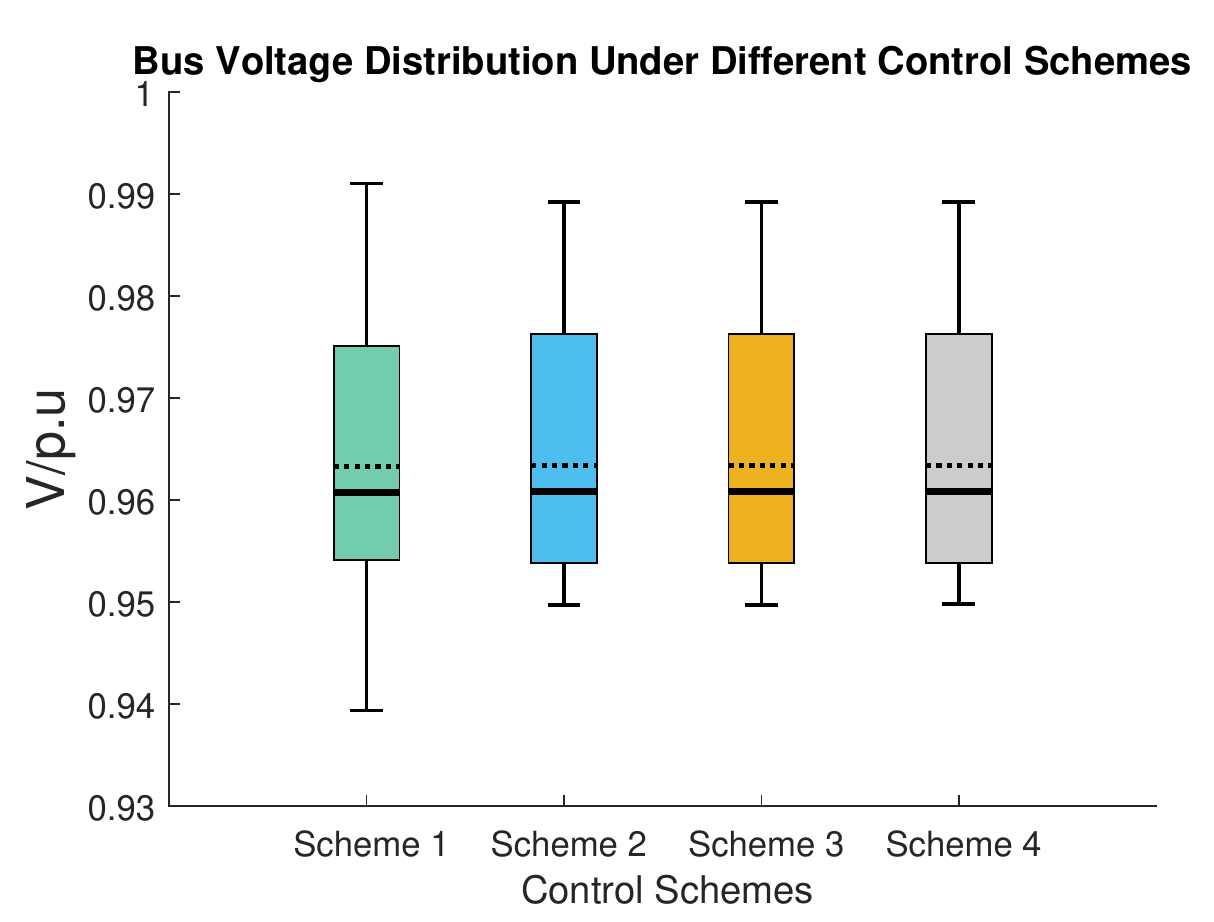}
    \caption{Distribution of system bus voltage under different control schemes}
    \label{Vbox}
\end{figure}

First, consider one extreme scenario, where all the PV generation is at the lowest level within the uncertainty set. The voltage profiles of the modified IEEE-123 bus test system under different schemes are presented in Fig.~\ref{V1point}, and the number of buses with voltage violations is given in Table.~\ref{NBViolation}. In Fig.~\ref{V1point},
the blue curves are the optimal voltage profiles determined in the first stage considering the forecast PV outputs, the yellow curves represent the voltage profiles in different schemes, and the red lines are voltage limits.
As shown in Fig.~\ref{V1point}, there are voltage violations for a considerable number of buses in Scheme 1. It indicates that without the second-stage reactive power adjustment, the first-stage VVC can not maintain the voltage profiles within the acceptable range. With respect to Scheme 2 and Scheme 3, both of them utilize the accurate voltage sensitivities.
The only difference between Scheme 2 and Scheme 3 is the implementation manner, where Scheme 2 is centralized and Scheme 3 is distributed. The outcomes for Scheme 2 and Scheme 3 are virtually identical, it validates our proposed distributed consensus-based AAARVC can converge to the optimal solution solved by the centralized optimization, but it is more scalable and practical. As shown in Table.~\ref{NBViolation}, there is only one bus with voltage violations for Scheme 1 and 2, where the lowest bus voltage magnitude for Scheme 2 and Scheme 3 is 0.949 p.u., which is very close to 0.95. For Scheme 4, its outcomes are very close to Scheme 2 and Scheme 3. The only minor difference is the number of buses with voltage violations is 2 for Scheme 4, slightly larger than Schemes 2 and 3. Such a minor difference might be caused by the error between the accurate and estimated voltage sensitives. The extreme scenario shows that the proposed data-driven AARVVC can achieve a great performance in terms of voltage regulation.
\begin{table}[htbp]
	\centering
	\caption{Number of Buses with Voltage Violations}
	\label{tab:1}  
	\begin{tabular}{c cccc}
		\hline\hline\noalign{\smallskip}
		Scheme & 1 & 2 & 3 & 4\\
		\noalign{\smallskip}\hline
		Bus with Voltage Violations & 75 & 1 & 1 & 2\\
		\noalign{\smallskip}\hline
		Lowest Voltage (p.u.) & 0.929 & 0.949 & 0.949 & 0.949\\
		\noalign{\smallskip}\hline
		\hline
	\end{tabular}
	\label{NBViolation}
\end{table}
To further explore the performance of our proposed data-driven AARVVC for voltage regulation, a Monte-Carlo simulation is carried out to randomly generate 1500 scenarios, where the PV active power output is uniformly sampled from its respective uncertainty interval. The distributions of bus voltage magnitudes under different control schemes are presented in Fig.~\ref{Vbox}. As can be seen in Fig.~\ref{Vbox}, under Scheme 1, voltages can not be maintained within the pre-defined range and the lowest voltage can be lower than 0.94. For other 3 schemes, voltages can  always be maintained within the acceptable level in most scenarios. Table.~\ref{percentage} provides the ratios of bus voltage violation under different schemes. Without the second-stage VVC, 7.73 \% buses are operated under voltage violations while the proposed data-driven AARVVC method can greatly decrease the ratio to around 0.5\%, which is very close to the optimal performance of Scheme 2 and Scheme 3. The lowest voltage for Scheme 4 is slightly lower than 0.95 p.u.. Note that Scheme 3 is also based on our proposed distributed consensus-based AARVVC. Scheme 3 and Scheme 4 are more scalable and require fewer computation burdens compared to Scheme 2. Even though the performance of Scheme 4 is slightly inferior to Scheme 3, it is more computationally efficient as it intelligently relies on the DNN to predict voltage sensitivities.
\begin{table}[htbp]
	\centering
	\caption{Percentage of Buses with Voltage Violations}
	\label{tab:1}  
	\begin{tabular}{c cccc}
		\hline\hline\noalign{\smallskip}
		Scheme & 1 & 2 & 3 & 4\\
		\noalign{\smallskip}\hline
		Bus Voltage Violation (\%) & 7.73 & 0.47 & 0.47 & 0.53\\
		\noalign{\smallskip}\hline
		Lowest Voltage (p.u.) & 0.939 & 0.949 & 0.949 & 0.949\\
		\noalign{\smallskip}\hline
		\hline
	\end{tabular}
	\label{percentage}
\end{table}
\section{Conclusion}
This paper introduces a data-driven AARVVC strategy for voltage regulation against PV uncertainty. The data-driven AARVVC strategy includes two parts: the data-driven voltage sensitivity estimation and the distributed consensus-based AARVVC, which are performed in a distributed manner with the estimated voltage sensitivities. The voltage sensitivities are efficiently predicted by the DNN with the operating statuses of selected buses as the input. The effectiveness and superiority of the proposed  data-driven AARVVC strategy are tested on the modified IEEE-123 bus test system. The results show it can accurately and efficiently estimate voltage sensitivities and achieve a good voltage regulation performance in a distributed consensus-based manner. In the future, we will take into account the network topology change.

\bibliography{ref} % your .bib file
\end{document}